\documentclass[10pt,aps,prl,twocolumn,superscriptaddress,groupedaddress]{revtex4-1}
\pdfoutput=1 
\usepackage{fancyhdr}
\usepackage{graphicx}
\usepackage{xspace}
\usepackage{rotating}
\usepackage[normalem]{ulem}
\usepackage{braket}
\usepackage{verbatim}
\usepackage{xcolor}
\usepackage{hyperref}
\usepackage[utf8]{inputenc}
\graphicspath{{./}{./Figs/}{./figs/}{../Figs/}}

\usepackage{amsfonts}
\usepackage{mathtools}
\usepackage{pifont}

\newcommand{\benum}{\begin{enumerate}}
\newcommand{\eenum}{\end{enumerate}}
\newcommand{\bi}{\begin{itemize}}
\newcommand{\ei}{\end{itemize}}

\newcommand{\mpl}{M_{\mathrm{Pl}}}

\newcommand{\be}{\begin{equation}}
\newcommand{\ee}{\end{equation}}

\newcommand{\nnmb}{\nonumber}
\newcommand{\bea}{\begin{eqnarray}}
\newcommand{\eea}{\end{eqnarray}}
\newcommand{\beq}{\begin{eqnarray}}
\newcommand{\eeq}{\end{eqnarray}}

\newcommand{\Rmnum}[1]{\expandafter\@slowromancap\romannumeral #1@}

\newcommand{\tv}{\bar{v}}
\newcommand{\lrf}[2]{\left(\frac{#1}{#2}\right)}

\def\gev{\,{\rm GeV}\,}

\begin{document} 
\title{
%The Return of Cosmic Strings through the Mist of Inflation and Gravitational Wave Bursts} 
Gravitational Wave Bursts as Harbingers of Cosmic Strings Diluted by Inflation}
%Observable Gravitational Wave Signatures from the Return of Pre-inflationary Cosmic Strings}

\author{Yanou Cui}
\email[]{yanou.cui@ucr.edu}
\affiliation{Department of Physics and Astronomy, University of California, Riverside, CA 92521, USA}
\author{Marek Lewicki}
\email[]{marek.lewicki@kcl.ac.uk}
\affiliation{Kings College London, Strand, London, WC2R 2LS, United Kingdom}
\affiliation{Faculty of Physics, University of Warsaw ul.\ Pasteura 5, 02-093 Warsaw, Poland}
\author{David E. Morrissey}
\email[]{dmorri@triumf.ca}
\affiliation{TRIUMF, 4004 Wesbrook Mall, Vancouver, BC, Canada V6T 2A3}
\date{\today}

\begin{abstract}
A standard expectation of primordial cosmological inflation is that it dilutes all relics created before its onset to unobservable levels.  We present a counterexample to this expectation by demonstrating that a network of cosmic strings diluted by inflation can regrow to a level that is potentially observable today in gravitational waves~(GWs).  In contrast to undiluted cosmic strings, whose primary GW signals are typically in the form of a stochastic GW background, the leading signal from a diluted cosmic string network can be distinctive bursts of GWs within the sensitivity reach of current and future GW observatories.  
%The discovery of such a signals may also provide insights on the inflationary epoch.%#

\end{abstract}
\maketitle
\section{Introduction} \label{sec:intro}

Cosmological observations provide strong support for initial conditions of the form expected from inflation followed by reheating: a hot thermal plasma that is very uniform with small, approximately adiabatic and scale-invariant density fluctuations over distances that are much larger than the Hubble length in that era~\cite{Aghanim:2018eyx,Akrami:2018odb}.  In addition to these features, inflation is expected to have diluted any relics created prior to its beginning, such as superheavy massive particles or most types of topological defects, to negligible levels~\cite{Guth:1980zm,Linde:1981mu,Albrecht:1982wi}.  In this work we investigate cosmic strings as a counterexample to these expectations, and we study the distinctive gravitational wave~(GW) signals they produce in this scenario.  An observation of such GW signals could provide new information about inflation or potential alternatives.

Cosmic strings are effectively one-dimensional objects that arise in
many theories of fundamental physics.  They can be fundamental objects~\cite{Copeland:2003bj,Dvali:2003zj,Polchinski:2004ia,Jackson:2004zg,Tye:2005fn} 
or configurations of quantum fields such as those originated from a $U(1)$ symmetry
breaking~\cite{Nielsen:1973cs,Kibble:1976sj}, 
but at macroscopic distances they are characterized almost completely 
by their energy per unit length (tension) $\mu$~\cite{Vilenkin:2000jqa}.  
In the early universe, cosmic strings are expected to form a network consisting 
of stable horizon-length long strings together with smaller closed loops that can decay away.  
The interplay between the slow stretching of long strings and the formation of new loops through
string intersections allows a cosmic string network to reach a scaling regime in which the total
energy density of the network tracks the dominant source of cosmological energy with a relative
fraction on the order of $G\mu$~\cite{Albrecht:1984xv,Bennett:1987vf,Allen:1990tv}, 
where $G = 1/8\pi\mpl^2$ is Newton's constant and $\mpl$ is the reduced Planck mass.

For cosmic strings created after inflation but well before today, 
scaling is predicted to be achieved reasonably soon after the formation of the network.  
To maintain scaling energy must be transferred from the network to radiation, 
and for local or fundamental strings this is usually expected to be in the form 
of gravitational waves emitted by closed string loops~\cite{Vilenkin:1981bx,Vachaspati:1984gt,Turok:1984cn,Burden:1985md,Olum:1999sg,Moore:2001px,Matsunami:2019fss} 
(although see Refs.~\cite{Vincent:1997cx,Bevis:2006mj,Figueroa:2012kw} that argue 
for mainly particle emission).  As a result, the most promising observational signal 
from such strings can be the stochastic gravitational wave background~(SGWB) 
created by the combined and unresolved emission of GWs by closed string loops over the history 
of the cosmos~\cite{Allen:1996vm,Blanco-Pillado:2013qja,Blanco-Pillado:2017rnf,Blanco-Pillado:2017oxo,Ringeval:2017eww,Abbott:2017mem,Caprini:2018mtu,Auclair:2019wcv,Auclair:2019jip}. 
The characteristic frequency spectrum produced this way could be also used to probe 
the expansion history of the universe~\cite{Cui:2017ufi,Cui:2018rwi,Chang:2019mza,Gouttenoire:2019rtn,Gouttenoire:2019kij}. 
In addition to the SGWB, more recent bursts of GWs from cusps or kinks on string loops 
can potentially be resolved as individual events~\cite{Damour:2000wa,Damour:2001bk,Damour:2004kw,Siemens:2006vk}, 
but for standard scaling strings they are harder to find than the unresolved SGWB~\cite{Abbott:2017mem,Blanco-Pillado:2017oxo,Auclair:2019wcv}.  

The situation can be much different for a cosmic string network formed before or shortly 
after the start of inflation.  Like any other relic, such strings would be exponentially diluted 
by the subsequent inflationary expansion.  However, following inflation the energy density in
long cosmic strings only falls off as $a^{-2}$, where $a(t)$ is cosmological scale factor, 
which is much slower than the $a^{-4}$ dilution of radiation and $a^{-3}$ of matter that 
are expected to dominate the energy budget of the cosmos until close to the present.  
This difference in dilution rates implies that a network of strings can potentially regrow 
and achieve scaling by today.  In such a string scenario, Ref.~\cite{Guedes:2018afo} showed 
that the SGWB is very strongly suppressed at moderate to high frequencies relative to a string 
network created after inflation, making it more difficult to observe.  For very strong inflationary
dilution, the SGWB can also be suppressed at the lower frequencies probed by pulsar timing arrays 
such as the PPTA~\cite{Shannon:2015ect} allowing for larger string tensions $G\mu$ to be consistent 
with existing bounds~\cite{Kamada:2012ag,Kamada:2014qta,Ringeval:2015ywa}.

In this \textit{Letter} we study the regrowth of cosmic strings after inflationary dilution 
and we investigate ways to discover them.  We demonstrate that individual resolved bursts 
of GWs can be the leading signal of such strings, and that such bursts could potentially 
be observed in the current LIGO experiment~\cite{TheLIGOScientific:2014jea,Thrane:2013oya,TheLIGOScientific:2016wyq}, 
or planned future experiments such as LISA~\cite{Bartolo:2016ami},
ET~\cite{Punturo:2010zz,Hild:2010id}, AION/MAGIS~\cite{Graham:2016plp,Graham:2017pmn,Badurina:2019hst}, 
and AEDGE~\cite{Bertoldi:2019tck}.  This is also true in cases where the SGWB is significantly 
suppressed at frequencies relevant to pulsar timing and thus reducing the limits from
PPTA~\cite{Shannon:2015ect} and the future reach of the planned SKA~\cite{Janssen:2014dka}.

\section{Inflation and String Regrowth}

To estimate the dilution of cosmic strings formed in the early stages of inflation and their
subsequent evolution, we use a simplified picture of inflation and reheating together with the
velocity-dependent one-scale~(VOS) model to describe the long-string network~\cite{Martins:1995tg,Martins:1996jp,Martins:2000cs,Avelino:2012qy,Sousa:2013aaa}. 
%We expect this treatment to capture the essential behaviour of more specific models of inflation and cosmic strings. 

During inflation we assume a constant Hubble parameter $H_I = V_{I}/3\mpl^2$ with $V_{I} \equiv M^4$
describing the inflationary energy density from initial time $t_I$ to end time $t_E$.  
Current observations limit $M \lesssim 10^{16}\,\gev$~\cite{Akrami:2018odb}
and we note that $GM^2 \simeq 7\times 10^{-7}$. 
The cosmological scale factor grows as $a(t)\propto e^{H_It}$.  
After inflation, we assume a reheating period that initiates a radiation-dominated 
phase with temperature  $T_{RH} \leq (30/\pi^2g_*)^{1/4}M$. 
We assume further that the temperature during reheating remains low enough to avoid 
the destruction of the pre-existing cosmic strings by thermal processes such as symmetry restoration.  

In the VOS model that we use to describe horizon-length long strings during 
and after inflation~\cite{Guedes:2018afo}, the long string energy is characterized by 
a correlation length parameter $L$ and velocity parameter $\tv$ such that 
the energy density of long strings is given by~\cite{Vilenkin:2000jqa}
\beq
\rho_{\infty} ~\equiv~ \frac{\mu}{L^2} \ .
\eeq
These parameters evolve according to~\cite{Martins:1996jp,Martins:2000cs}
\beq
\frac{dL}{dt} &=& (1+\tv^2)\,HL + \frac{\tilde{c}\tv}{2}
\label{eq:vos1}\\
\frac{d\tv}{dt} &=& (1-\tv^2)
\left[\frac{k(\tv)}{L} - 2H\,\tv\right] \ ,
\label{eq:vos2}
\eeq
where 
\beq
k(\tv) = \frac{2\sqrt{2}}{\pi}(1-\tv^2)(1+2\sqrt{2}\tv^3)
\left(\frac{1-8\tv^6}{1+8\tv^6}\right) \ ,
\eeq
and $\tilde{c} \simeq 0.23$ describes closed loop formation~\cite{Martins:2000cs}.

As initial condition, we take
\beq
L(t_F) \equiv L_F = \frac{1}{{\zeta}\,H_I} \ ,
\eeq
where $t_F$ is the greater of the beginning of inflation or the network
formation time, and ${\zeta}^2$ corresponds approximately to the number 
of long strings within the Hubble volume at time $t_F$.
After $t_F$, the string network parameters quickly reach 
an attractor solution during inflation  
(independent of the value of $\tv$ at $t_F$) given by
\beq
L(t) = L_F\,e^{H_I(t-t_F)} \ ,~~~\tv(t) = \frac{2\sqrt{2}}{\pi}\frac{1}{H_IL(t)} \ .
\eeq
This solution reflects the dilution of the long string network,
with $HL \gg 1$ and $\tv \ll 1$ by the end of inflation.  While $HL \gg 1$ after inflation, 
The subsequent evolution of the string network after inflation takes a very simple form
while $HL \gg 1$ with $(L/a)$ approximately constant.  It follows that $HL$ decreases
after inflation, corresponding to the gradual regrowth of the string network.
If the network is to produce a potentially observable signal in GWs, 
at least a few strings are needed within our current Hubble volume corresponding
to $HL \lesssim 1$ today.

To determine the conditions under which there is enough string 
regrowth for $HL \to 1$ while also maintaining a sufficient amount of 
inflation, it is useful to compare the evolution of $L$ prior to 
scaling to that of the curvature radius $R = 1/(H\sqrt{|\Omega\!-\!1|})$
which evolves in precisely the same way, independently of the details 
of inflation or reheating. In the absence of strong tuning of the 
curvature radius at the start of inflation, the current $95\%$ limit 
on curvature $|\Omega_0\!-\!1| = 0.0007\pm 0.0037$~\cite{Aghanim:2018eyx} 
puts the strongest  lower bound on the total number of $e$-foldings 
of inflationary expansion~\cite{Akrami:2018odb}.  
Defining $\Delta N \geq 0$ to be the number of $e$-foldings between $t_I$ and $t_F$, 
the  total number of inflationary $e$-foldings is 
$N_{tot} = H_I(t_E-t_I) \equiv N_F+\Delta N$. 
Note that $\Delta N=0$ corresponds to the string forming before or at the start of inflation. 
Applying the curvature limit on $N_{tot}$, we find
\beq
\Delta N + \ln\zeta &~\geq~&
2.7  + \frac{1}{2}\ln(|\Omega_{I}\!-\!1|) 
\label{eq:zscl}\\ 
&&
{}\hspace{-1.7cm}+ \frac{1}{2}\ln\left[
\Omega_{\Lambda}(1+\tilde{z})^{-2}
+ \Omega_m(1+\tilde{z})
+ \Omega_r(1+\tilde{z})^2
\right] 
\nnmb
\eeq
where $\tilde{z}$ is the redshift at which $HL\to 1$, $\Omega_a$ 
are the fractional energy densities in dark energy, matter, and radiation 
relative to critical today, 
$|\Omega_I\!-\!1|$ is the deviation from flatness at the start of inflation,
$\ln(|\Omega_0-1|)/2 \leq 2.7$ is the bound from Planck~\cite{Aghanim:2018eyx}, 
and the last line describes the additional evolution of $(\Omega-1)$ 
between $\tilde{z}$ and now.

% \begin{figure}[ttt]
% \centering
% \includegraphics[width=0.48\textwidth]{vosscl.pdf}
% \caption{Minimum value of $(\Delta N + \ln\zeta)$ needed for the network to reach $HL\to 1$ by redshift $\tilde{z}$ for $|\Omega_I\!-\!1| = 1,\,0.1,\,0.01$.
% \label{fig:vosscl}}
% \end{figure}

% In Fig.~\ref{fig:vosscl} we show the minimal value of $\Delta N + \ln\zeta$ needed for the network 
% to reach $HL=1$ by redshift $\tilde{z}$ for three values of the curvature at the start of inflation: 
% $|\Omega_I\!-\!1| = 1,\,0.1,\,0.01$.  
% The flattening of the curves at low $\tilde{z}$ reflects the recent onset of dark energy domination.
% This figure shows that $HL \to 1$ requires that the network must either have been formed 
% $\Delta N$ $e$-folds after the start of inflation, or the initial number of long strings within 
% the Hubble volume at time $t_I$ must have been much greater than unity.

To fulfill this bound the network must either have been formed $\Delta N$ $e$-folds after the start of inflation, or the initial number of long strings within 
the Hubble volume at time $t_I$ must have been much greater than unity.
One or both of these features can arise if the string network was formed 
in a (weakly) first-order phase transition.  
Such a transition can produce an initial 
correlation length much shorter than the Hubble radius, 
potentially yielding many strings within the Hubble volume at formation and thus 
$\zeta \gg 1$~\cite{Rajantie:2001ps}.  Furthermore, a strong super-cooled 
transition can be initially slow relative to $t_I \sim 1/H_I$, 
in which case it will not complete until after several $e$-folds of expansion 
to yield $\Delta N > 0$~\cite{Turner:1992tz,Konstandin:2011dr,Ellis:2018mja,Ellis:2019oqb}.
We defer a detailed study of the realization of such conditions 
within specific theories of inflation and string formation to a future work, 
although we note that the supercooled-transition scenario of
Ref.~\cite{Konstandin:2011dr} finds values as large as $\Delta N \sim 18$
as well as the previous studies of string creation in the early stages of inflation 
of Refs.~\cite{Shafi:1984tt,Vishniac:1986sk,Yokoyama:1989pa,Hodges:1991xs,Basu:1991ig,Kamada:2012ag,Kamada:2014qta}.

\section{String Scaling and Loop Formation}

\begin{figure}
\centering
\includegraphics[width=0.45\textwidth]{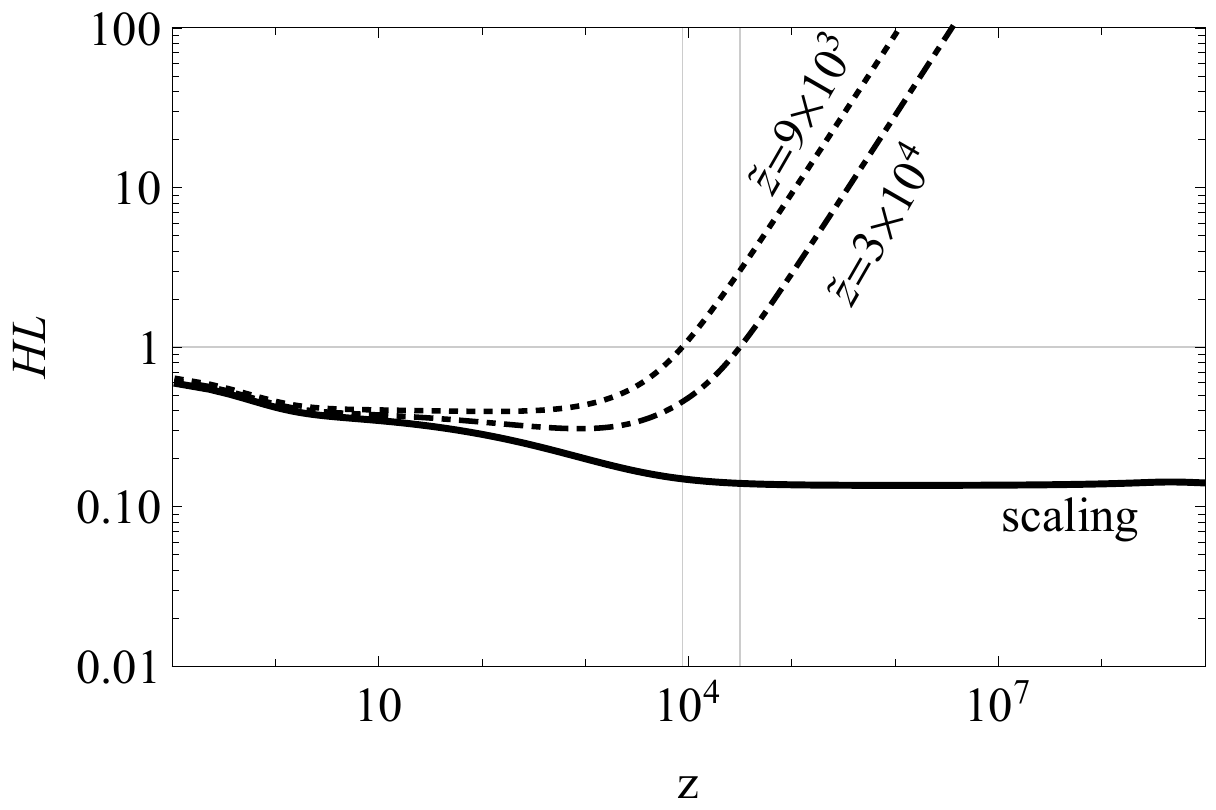}
\caption{Evolution of the VOS model string length parameter $HL$ 
for diluted strings compared to that of a scaling string network 
for two representative values of the transition redshift 
$\tilde{z} = 9\times 10^{3},\,3\times 10^{4}$ at which $HL\to 1$.
\label{fig:LHplot}}
\end{figure}

Once $HL$ approaches unity at redshift $\tilde{z}$, the VOS string length parameter 
$L$ begins to deviate from the simple $L\propto a$ form and evolves together 
with $\tv$ toward the string scaling limit.  
In general, we find that the network does not reach
scaling until considerably later than $\tilde{z}$.  This is shown in 
Fig.~\ref{fig:LHplot} for two representative values of $\tilde{z} = 
9\times 10^{3},\,3\times 10^{4}$.  Using Eq.~\eqref{eq:zscl}, 
we find that these values of $\tilde{z}$ correspond to 
$\Delta N + \ln\zeta \geq 7.3$ and $8.4$ 
which are plausible in the context of supercooled scenarios 
such as that of Ref.~\cite{Konstandin:2011dr}.

To determine the density distribution of closed string loops, we use 
the VOS model described above to find the total rate of loop 
production prior to and during the scaling regime together with the 
results of simulations to estimate their initial size.  Recent 
simulations find that (Nambu-Goto) string scaling networks produce a 
population of larger loops that are moderately non-relativistic with 
initial size $l_i = \alpha\,L(t_i)$ with $\alpha \sim 0.1$ as well as 
a collection of smaller loops that are highly 
relativistic~\cite{Blanco-Pillado:2013qja}. Most of the energy 
transferred to the smaller loops is in the form of kinetic energy that
simply redshifts away, and thus larger loops are expected to be the 
dominant source of GWs and we focus exclusively on them.  The large 
loops are found to make up a fraction $\mathcal{F}_{\alpha} \sim 0.1$ 
of the total energy transferred to 
loops~\cite{Blanco-Pillado:2013qja,Blanco-Pillado:2017oxo}.  After 
formation, these large loops oscillate, emit energy in the form of 
GWs, and gradually shorten according to
\beq
l(t) = \alpha\,L(t_i) - \Gamma\,G\mu\,(t-t_i) \ ,
\label{eq:ltime}
\eeq
where $\Gamma \simeq 50$ is the total rate of GW 
emission~\cite{Burden:1985md,Vilenkin:2000jqa,Blanco-Pillado:2017oxo}.  
Matching these simulation results to the total rate of energy loss to 
loop formation within the VOS model, corresponding to the 
$\tilde{c}\tv/2$ term in Eq.~\eqref{eq:vos1}, the differential number 
density of long loops per unit length 
is~\cite{Sousa:2013aaa,Auclair:2019wcv}
\beq
n(l,t) \equiv 
\frac{\mathcal{F}_{\alpha}}{\sqrt{2}}\; %\left(\frac{a(t_i)}{a(t)}\right)^3
%\frac{ \left(a(t_i)/a(t)\right)^3}
%\left(\frac{z(t)+1}{z(t_i)+1}\right)^3
%\frac{ \left(a(t_i)/a(t)\right)^3}
\frac{ \left[z(t)+1\right]^3/\left[z(t_i)+1\right]^3}
{\alpha\,dL/dt|_{t=t_i}+\Gamma\,G\mu}\;
\frac{\tilde{c}\,\tv(t_i)}{\alpha\,L^4(t_i)} \ ,
\eeq
where $t_i$ on the right hand side is to be determined in terms of $l$
and $t$ through Eq.~\eqref{eq:ltime}.
% In Fig.~\ref{fig:nplot} we show the differential density of loops 
% $n(l,t)$ as a function of redshift from three loop sizes $l = 
% \alpha\,L(t)$~(left), $l=\alpha\, L(t)/10$~(middle), and $l=\alpha\, 
% L(t)/100$~(right) for an undiluted string network in the scaling 
% regime as well as for diluted networks with $\tilde{z} = 9\times 
% 10^{3},\,3\times 10^{4}$.  The loop density is clearly suppressed at 
% redshifts $z > \tilde{z}$.

% %%%%%%%%%%%%%%%%%%%%%%%%%%%%%%%%%%%%%%%%%%%%%%%%%%%%%%%%%%%%%%%%%%%%%%%%%%%%%%%%%
% \begin{figure*}
% \centering
% \includegraphics[height=5.0cm]{nplot1.pdf}
% \includegraphics[height=5.0cm]{nplot2.pdf}
% \includegraphics[height=5.0cm]{nplot3.pdf}
% \caption{
% Differential closed string loop density $t^4\,n(l,t)$ for specific loop sizes $l = \alpha L(t)$~(left), $l=\alpha\, L(t)/10$~(middle), and $l = \alpha\, L(t)/100$~(right) with $\alpha = 0.1$ as a function of redshift $z$.  The solid black line shows the density expected from an undiluted string network in the scaling regime, which the dashed and dash-dotted lines show the densities for diluted string networks with $\tilde{z} = 9\times 10^{3},\,3\times 10^{4}$.
% \label{fig:nplot}}
% \end{figure*}
%%%%%%%%%%%%%%%%%%%%%%%%%%%%%%%%%%%%%%%%%%%%%%%%%%%%%%%%%%%%%%%%%%%%%%%%%%%%%%%%%%

\section{Gravitational Wave Signals from String Loops}

String loops oscillate and lose energy to GWs.  Much of the emission 
to GWs comes from short, violent, collimated bursts involving 
cusps or kinks on the string loops.  Bursts emitted by a cosmic string
network over its cosmic history that are not resolved contribute to 
the characteristic stochastic gravitational wave background~(SGWB) of the
network.  More recent bursts can also potentially be observed as 
distinct, individual events. In this section we calculate the GW burst
signal following the methods of 
Refs.~\cite{Damour:2000wa,Damour:2001bk,Damour:2004kw} together with 
the refinements of Refs.~\cite{Siemens:2006vk,Abbott:2017mem,Auclair:2019wcv}.

The GWs produced by a burst from a cusp or a kink are mostly collimated within a beaming angle
\bea \label{eq:beaming}
\theta_m(l,z,f)=\left[(1+z)f l\right]^{-\frac{1}{3}} < 1 \ ,
\eeq
where $f$ is the GW frequency seen today and $z$ is the redshift at emission.
Within this angular region, 
the GW waveform is~\cite{Damour:2000wa,Damour:2001bk,Damour:2004kw}
\beq \label{eq:waveform}
%h(l,a,f)=\frac{G\mu \,  l^{2-q}  }{a^{1-q} \, r(a)}f^{-q} \, , \ \  
h(l,z,f)=\frac{f^{-q}\,l^{2-q}}{(1+z)^{q-1}}\frac{G\mu}{r(z)}  \ , \ \  
r(z)=\int_0^z \!\frac{dz'}{H(z')}
\eeq
where $r(z)$ is the proper distance to the source
and $q=4/3 \, (5/3)$ for cusps (kinks).
%In principle both formulae above also contain a numerical factor taking into account various uncertainties in the calculation, however, as usually done in the literature we will set bot of them to one.

%The number density of cusp or kink features per unit string loop
%length per oscillation period is~\cite{Siemens:2006vk}
The rate of cusp or kink features per unit volume per unit string length
at emission is~\cite{Siemens:2006vk}
\be
\nu(l,z) = \frac{2}{l}\,N_q\,n(l,z) \ ,
\ee
where $N_q$ is the number of features within an oscillation period
and $n(l,z)$ is the loop density obtained above.
Putting these pieces together, the rate of bursts per volume 
per length observed today is~\cite{Abbott:2017mem}
\beq \label{eq:dRdVdl}
\frac{d^2R}{dV dl}(l,a,f) &=&
\\&&{}\hspace{-1.8cm} 
\frac{\nu(l,z)}{(1+z)} \left[ \frac{\theta_m(l,z,f)}{2} \right]^{3(2-q)}\Theta(1-\theta_m) \ .
\nnmb
\eeq
It is convenient to use 
\be
dV=  \frac{4\pi r^2(z)}{(1+z)^3H(z)}dz \ ,
\ee
and Eq.~\eqref{eq:waveform} to rewrite Eq.~\eqref{eq:dRdVdl} as
\beq
\frac{d^2R}{dz\,dh} &=&
%\\&&{}\hspace{-1.8cm}
\frac{2^{3(q-1)}\,\pi\, G\mu\,  N_q}{(2-q)}
\frac{r(z)}{(1+z)^5H(z)}\,
\frac{n(l,z)}{h^2f^2}
%n(\left(\frac{h\, f^q\, r(a)}{G\mu\, a^{q-1}}\right)^{\frac{1}{2-q}},t(a)).
%\nnmb
\eeq
where $l$ is now a function of $h$, $f$, and $z$, and the consistency constraints
$\theta_m < 1$ and $l< \alpha\,L(z)$ are enforced by
restricting $h\in [h_{min},\,h_{max}]$ with
\be
h_{\rm min}=\frac{1}{(1+z)f^2}\frac{G\mu}{r(z)} \ ,~
h_{\rm max}=\frac{[\alpha L(z)]^{q-2}}{f^q(1+z)^{q-1}} 
\frac{\, G\mu}{r(z)} 
\ee
Since the total GW signal is expected to be dominated by bursts from
cusps~\cite{Ringeval:2017eww,Blanco-Pillado:2017oxo}, 
we set $q=4/3$ and $N_q=2.13$ (to match $\Gamma = 50$~\cite{Auclair:2019wcv})
in the analysis to follow.  

To compare these GW signals with current and future detectors, we 
separate the contributions into more recent bursts of large amplitude 
that can be resolved individually from earlier ones that are not 
resolved and contribute to the net stochastic background.  
If a burst is to be resolved in a given frequency band $f$, 
it must produce a strain greater than the experimental sensitivity $h > h_{\rm exp}$ with rate less than $f$.  
The rate of such events is~\cite{Siemens:2006vk,Auclair:2019wcv}
\beq
R_{\rm exp}(f) = \int_{0}^{z_*} \!dz \int_{\max(h_{\rm min},h_{\rm exp})}^{h_{\rm max}} 
\!\!\!dh \; \frac{d^2R}{dz\, dh}(h,z,f)~~~~~ \ 
\eeq
where $z_*$ enforces the rate condition and is given by
\be \label{eq:astar}
f= \int_{0}^{z_*}\!dz \int_{h_{\rm min}}^{h_{\rm max}}\!dh \;  
\frac{d^2R}{dz\, dh}(h,z,f) \, .
\ee
Unresolved bursts contribute to the SGWB as~\cite{Siemens:2006vk,Auclair:2019wcv}
\be \label{eq:OmegaGW}
\Omega_{\rm GW}(f) =\frac{4\pi^2 f^3}{3H_0^2}
\int_{z_*}^{\infty}\!dz\,\int_{h_{\rm min}}^{h_{\rm max}}\!dh \; 
h^2\,%\frac{f^3h^2}{\rho_c} \,
\frac{d^2R}{dz\, dh}(h,z,f) 
\ee
with $\Omega_{GW} = (f/\rho_c)\,d\rho_{\rm GW}/df$
for critical density $\rho_c$. 

\begin{figure}
\centering
\includegraphics[width=0.48\textwidth]{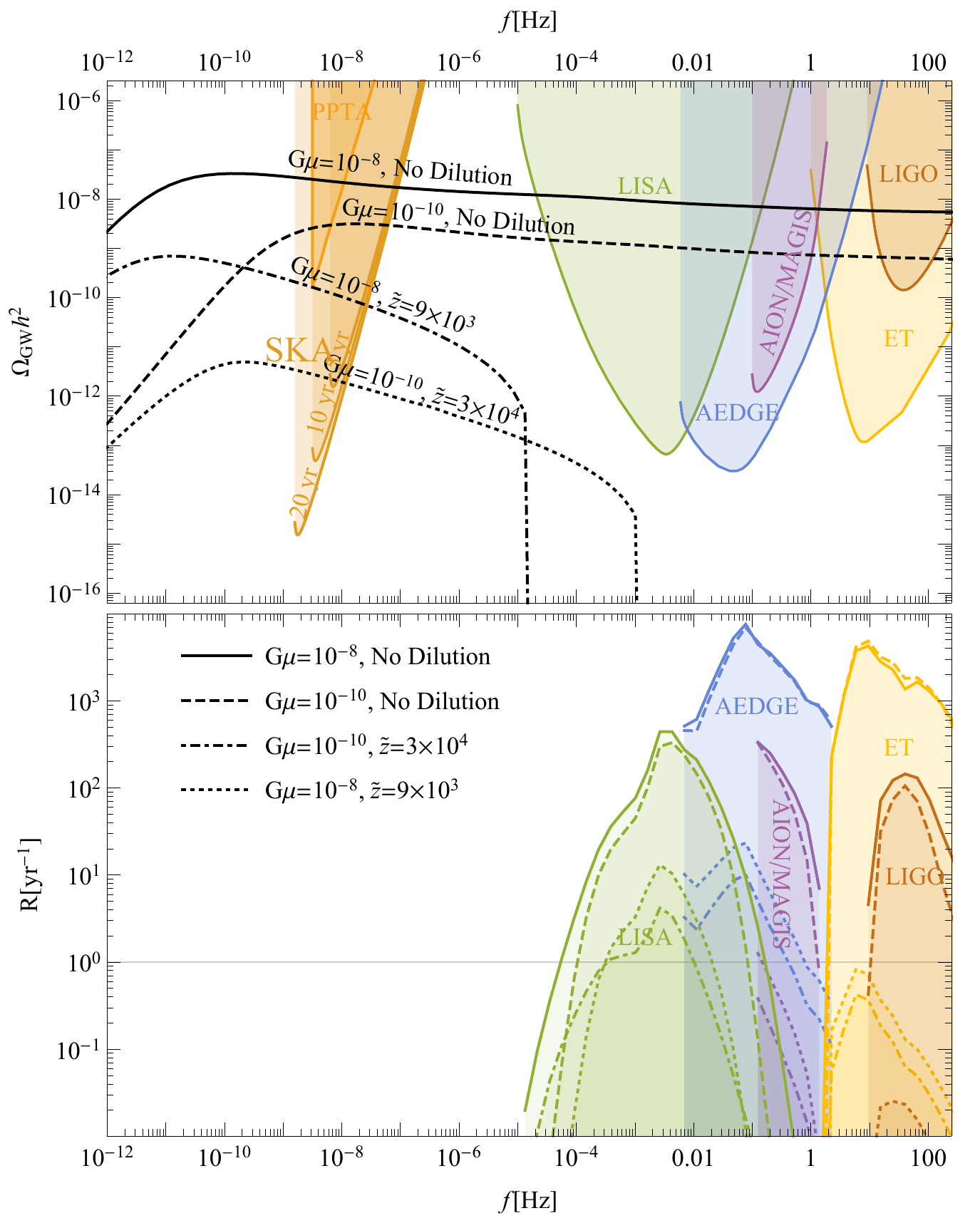}
\caption{
Gravitational wave signals from diluted and undiluted cosmic string networks
as a function of frequency observed today together with the corresponding
sensitivities of the indicated observatories.  The top panel shows the
stochastic GW background while the lower panel gives the event rates of
resolved bursts.  In both panels we show curves for $G\mu =
10^{-8},\,10^{-10}$ for undiluted networks as well as for two diluted
networks with $\tilde{z} = 9\times 10^{3}$ for $G\mu = 10^{-8}$, 
and $\tilde{z} = 3\times 10^{-4}$ for $G\mu = 10^{-10}$.
\label{fig:StringRegrowthplot}
}
\end{figure}

In Fig.~\ref{fig:StringRegrowthplot} we show the SGWB from unresolved
bursts~(top panel) as well as the resolved burst rate as a function of
frequency~(bottom panel) for diluted and undiluted cosmic string networks.  
% In both scenarios we show curves for $G\mu = 10^{-8},\,10^{-10}$, while for 
% the diluted networks we consider the representative values
% $\tilde{z} = 9\times 10^3$ for $G\mu = 10^{-8}$, and $\tilde{z} = 3\times 10^4$ for $G\mu=10^{-10}$. 
Also shown in these figures are 
the expected sensitivity ranges of various GW observatories including the 
current LIGO~\cite{TheLIGOScientific:2016wyq}, and planned 
LISA~\cite{Bartolo:2016ami}, ET~\cite{Punturo:2010zz,Hild:2010id}, 
AION/MAGIS~\cite{Graham:2016plp,Graham:2017pmn,Badurina:2019hst}, and 
AEDGE~\cite{Bertoldi:2019tck}, as well as the existing 
PPTA~\cite{Shannon:2015ect} and planned SKA~\cite{Janssen:2014dka} 
pulsar timing arrays.
The top panel of Fig.~\ref{fig:StringRegrowthplot} shows that undiluted
string networks with $G\mu=10^{-8},\,10^{-10}$ are already excluded 
by pulsar timing measurements at the PPTA~\cite{Shannon:2015ect},
but they can be consistent with diluted strings~\cite{Kamada:2012ag}.  Despite this strong suppression of the SGWB, the lower panel of 
Fig.~\ref{fig:StringRegrowthplot} shows that resolved burst events
from diluted string networks could still be seen in future 
gravitational wave observatories.  %The resolved GW burst signal can therefore be the leading discovery channel for diluted cosmic strings at gravitational wave observatories.

We note that the SGWB for the diluted networks computed using the burst
method described above falls off as $f^{-1/3}$ at high frequency, 
in contrast to the $f^{-1}$ behavior found in Refs.~\cite{Guedes:2018afo,Gouttenoire:2019kij}
computed in a different way by summing over averaged loop normal mode emissions.  
These two methods predict very similar SGWB signals from 
cosmic strings in the scaling regime~\cite{Abbott:2017mem,Auclair:2019wcv}.
This discrepancy appears to be the result of not including a sufficient number 
of modes when computing the SGWB with the normal mode method. 
In the supplemental material below%~\footnote{The supplemental material includes references~\cite{Sanidas:2012ee,Blanco-Pillado:2015ana,Sousa:2020sxs,Blasi:2020wpy}.}
, we show that the number of normal modes needed 
to compute the SGWB for a diluted cosmic string network can be orders of magnitude larger than for a cosmic string that reaches scaling very early,
and that when a large number is required the signal computed with 
the normal mode method asymptotes to $\Omega_{GW} \propto f^{-1/3}$, 
in agreement with the burst method used in this work.
At higher frequencies, we also find a sharper drop from the burst 
method caused by the subtraction of infrequent bursts from the SGWB.

%Beyond that, the very fast fall off that the diluted SGWB exhibit at very high frequency is caused by subtraction of the infrequent bursts. Essentially, for diluted networks at very high frequencies the total rate drops below the measurement frequency (see Eq.\eqref{eq:astar}) and all the bursts are rare enough to be observed individually.

%%%%%%%%%%%%%%%%%%%%%%%%%%%%%%%%%%%%%%%%%%%%%%%%%%%%%%%%%%%%%%%%%%%%%%%%%%%%%%%%%%
\begin{figure}
\centering
\includegraphics[width=0.48\textwidth]{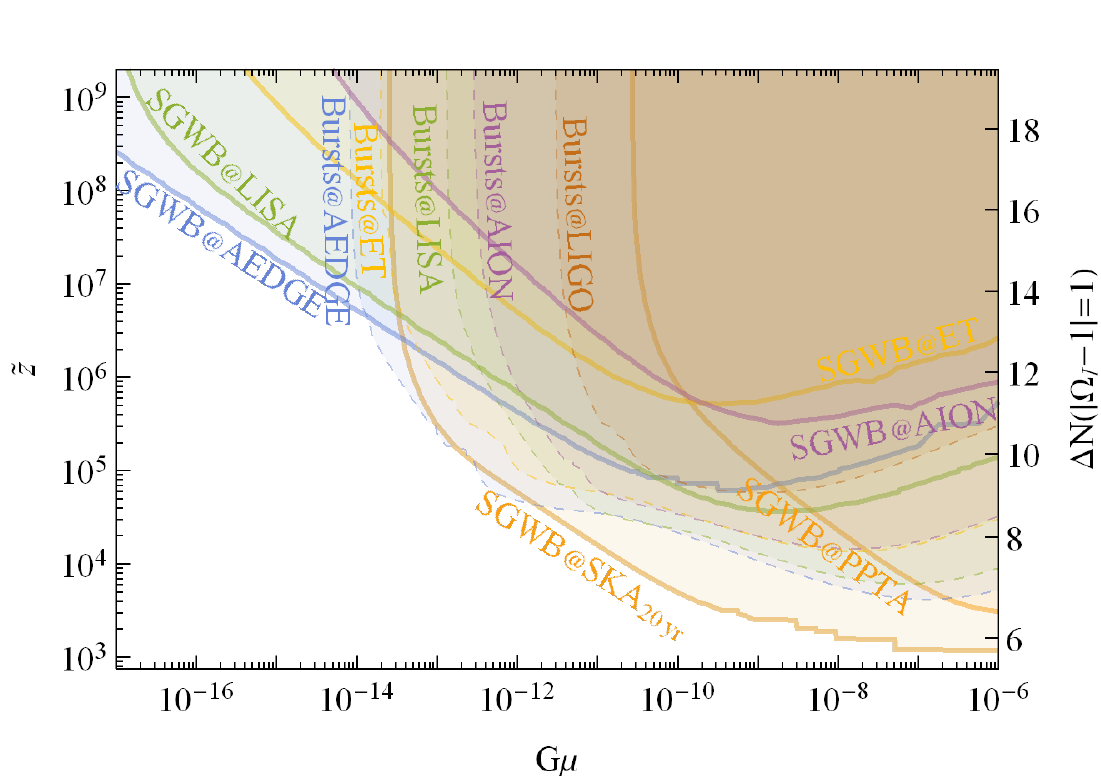}
\caption{
Sensitivity of current and future GW experiments to signals from diluted cosmic strings as a function of tension $G\mu$ and the redshift $\tilde{z}$ at which loop production resumes. Solid lines show the sensitivity of the indicated experiments to stochastic backgrounds (with $\mathrm{SNR} \geq 10$) while dashed lines their sensitivity to burst signals (with rate $R\geq 1/\mathrm{yr}$). 
}
\label{fig:GWexpreachplot}
\end{figure}

%%%%%%%%%%%%%%%%%%%%%%%%%%%%%%%%%%%%%%%%%%%%%%%%%%%%%%%%%%%%%%%%%%%%%%%%%%%%%%%%%%
\indent To illustrate the potential observability of diluted strings for more 
general scenarios, we show in Fig.~\ref{fig:GWexpreachplot} the reach 
of current and future GW with related detectors as a function of the string
tension $G\mu$ and redshift $\tilde{z}$ when strings grow back into the
horizon.  The figure demonstrates that bursts can be the leading discovery
channel for a broad range of string tensions and dilution factors, with the
SGWB signal becoming more prominent for larger $\tilde{z}$ corresponding to
less dilution.  For very strong dilution with $\tilde{z} \lesssim 10^3$, the
string network may not have reached scaling even by the present time and the
combined GW signal is too weak to be observed in the foreseeable future. 
With such a dilution, other direct bounds such as the CMB distortion limit of
$G\mu < 1.1\times 10^{-7}$~\cite{Charnock:2016nzm} are expected to be
mitigated as well, although such a low $\tilde{z}$ regime may still be
observable in late-time astrophysical effects such as gravitational lensing
and imprints on structure formation~\cite{Sazhin:2006kf,Morganson:2009yk,Ringeval:2015ywa,Laliberte:2018ina,Fernandez:2020vgi}. 

%%%%%%%%%%%%%%%%%%%%%%%%%%%%%%%%%%%%%%%%%%%%%%%%%%%%%%%%%%%%%%%%%%%%%%%%%%%%%%%%%%
%%%%%%%%%%%%%%%%%%%%%%%%%%%%%%%%%%%%%%%%%%%%%%%%%%%%%%%%%%%%%%%%%%%%%%%%%%%%%%%%%%

\section{Conclusions}
In this \textit{Letter} we have demonstrated that a network of cosmic strings formed before or in the early stages of primordial inflation can regrow to observable levels today.  Furthermore, we have shown that in this scenario the stochastic gravitational wave signal is strongly suppressed and the most promising discovery channel can be distinctive burst signals in gravitational wave detectors.

{\it note added:}
 Our results recently gained also experimental relevance as NANOGrav Collaboration reported strong evidence for a stochastic common-spectrum process~\cite{Arzoumanian:2020vkk} , which can be interpreted as SGWB from cosmic strings~\cite{Ellis:2020ena,Blasi:2020mfx}. In fact our current results have a slight advantage over the vanilla cosmic string case as the current data favours a negative slope for the GW abundance which we generically obtain in our scenario at PTA frequencies. We have in fact verified that our first benchmark point $G\mu=10^{-8}\, ,  \tilde{z}=9\times10^{3}$ fits this new data at $95\%$ CL and a stronger signal would improve the fit.

\begin{acknowledgments}
\section*{Acknowledgements}
We thank 
Jose Juan Blanco-Pillado,
Djuna Croon,
James Wells,
and Graham White
for helpful discussions. 
YC is supported in part by the US Department of Energy under award number DE-SC0008541. 
ML is supported by the UK STFC Grant ST/P000258/1 and the Polish National Science Center grant 2018/31/D/ST2/02048.
DEM is supported by a Discovery Grant from the Natural Sciences and Engineering Research Council of Canada (NSERC), and TRIUMF, which receives federal funding via a contribution agreement with the National Research Council of Canada (NRC).
\end{acknowledgments}

%%%%%%%%%%%%%%%%%%%%%%%%%%%%%%%%%%%%%%%%%%%%%%%%%%%%%%%%%%%%%%%%%%%%%%%%%%%%%%%%%%%%%%%%%%%%%%%%%%%%
%%%%%%%%%%%%%%%%%%%%%%%%%%%%%%%%%%%%%%%%%%%%%%%%%%%%%%%%%%%%%%%%%%%%%%%%%%%%%%%%%%%%%%%%%%%%%%%%%%%%
\section*{Supplementary Material} \label{sec:sm}
In the main text we showed that the stochastic gravitational wave background~(SGWB) from a cosmic string network is suppressed at higher frequencies if the network only approaches the scaling regime density from below at a relatively late time in the history of the cosmos.  By summing the gravitational wave contributions of unresolved string loop bursts, we found a fall-off of the SGWB with frequency $f$ according to $\Omega_{GW} \propto f^{-1/3}$. This dependence differs from the $\Omega_{GW} \propto f^{-1}$ scaling found in Refs.~\cite{Guedes:2018afo,Gouttenoire:2019kij,Gouttenoire:2019rtn} computed by summing over string loop normal modes. 

Since both methods for computing the SGWB give consistent results for strings networks that reached the scaling regime at very early cosmological times~\cite{Abbott:2017mem,Auclair:2019wcv}, the origin of this disagreement is puzzling. We argue here that the late-scaling scenario requires summing over a much larger number of normal modes than is needed in the standard early scaling case.  When a sufficient number of modes are included, we show that the normal method predicts a regime of large-$f$ scaling $\Omega_{GW} \propto f^{-1/3}$, in agreement with the burst method. 

Beyond this specific scenario, we show further that our result also applies to a range of scenarios where the SGWB spectrum from cosmic strings is modified by a period of non-standard cosmological evolution. In particular, this range includes an early period of matter domination, as studied in this context in Refs.~\cite{Cui:2017ufi,Cui:2018rwi,Gouttenoire:2019kij,Gouttenoire:2019rtn}.

After this introduction, we review the normal mode method for computing the SGWB from a scaling string network. Next, we investigate the number of modes required for this approach to provide an accurate description of the full SGWB spectrum using a simplified model for late scaling, and we show how our results can be generalized to other scenarios. After this, we present a simple method to include an arbitrarily large number of modes for general string network histories. Finally, we show how the normal mode method and our result can be applied to cosmic string loop emission dominated by kinks or other features.

\subsection*{Review of the Normal Mode Method}

In the normal mode method, the GW emission from a cosmic string loop is treated as a sum over normal modes.  The frequency of emission from a cosmic string loop of length $l$ is given by~\cite{Auclair:2019wcv}
\beq
f_e  = \frac{2\pi\,k}{l} \ ,
\eeq
where $k \in \mathbb{Z}^+$ is the mode number. Each loop emits with total rate factor $\Gamma\simeq 50$ such that~\cite{Vilenkin:1981bx}
\beq
\frac{dl}{dt} = -\Gamma\,G\mu^2 \ .
\eeq
Simulations find $\Gamma \simeq 50$ with power per mode scaling with mode number as~\cite{Blanco-Pillado:2017oxo}
\beq
\Gamma = \sum_k\Gamma^{(k)} \ ,~~~~~
\Gamma^{(k)} \simeq \lrf{\Gamma}{3.60}k^{-4/3} \ .
\label{eq:pmode}
\eeq
The $k^{-4/3}$ scaling reflects the expectation that this emission is dominated by cusps.

The SGWB from a cosmic string network is obtained in this approach by summing over all loop emissions while including the appropriate redshift factors. Loops are created by the chopping of horizon-length long strings. In the scaling regime, detailed simulations find that a fraction $\mathcal{F} \simeq 0.1$ of the energy transferred from long strings to loops goes to large loops of initial size $l_i = \alpha\,t_i$ with $\alpha \simeq 0.1$, with the remaining energy transferred to highly relativistic smaller loops~\cite{Blanco-Pillado:2013qja,Blanco-Pillado:2015ana,Blanco-Pillado:2017oxo}. The long loops dominate the GW signal, and they are created at the rate
\beq
\frac{dn_\alpha}{dt} = \mathcal{F}\,\frac{C_{eff}}{\alpha}{t_i}^{-4} \ ,
\eeq
where $C_{eff}$ is on the order of unity. For a scaling string network formed at time $t_F$, the resulting SGWB at frequency $f$ seen today is~\cite{Auclair:2019wcv}
\beq
\Omega_{GW}(f) = \sum_{k=1}^{\infty}\Omega_{GW}^{(k)}(f) \ ,
\eeq
with
\beq
\Omega_{GW}^{(k)}(f) &=& \frac{1}{\rho_c}\frac{2k}{f}\frac{\mathcal{F}_{\alpha}\Gamma^{(k)}G\mu^2}{\alpha(\alpha+\Gamma G\mu)}\,
\label{eq:mode1}
\\
&&\int_{t_i}^{t_0}\!\!dt\;
\frac{C_{eff}}{t_i^4}
\lrf{a}{a_0}^5
\lrf{a_i}{a}^3
\Theta(t_i-t_F) \ ,
\nnmb
\eeq
where the integral runs over the loop emission time $t$ and $t_i$ is the corresponding loop formation time given by
\beq
t_i(t,f;k) = \frac{1}{\alpha+\Gamma G\mu}\left[
\frac{2k}{f}\lrf{a}{a_0} + \Gamma G\mu\,t
\right] \ .
\label{eq:ti}
\eeq
A useful observation is that 
\beq
\Omega^{(k)}_{GW}(f) = k^{-4/3}\,\Omega_{GW}^{(1)}(f/k) \ .
\label{eq:gwkscl}
\eeq
We will use this extensively below.

%%%%%%%%%%%%%%%%%%%%%%
\begin{comment}
An equivalent (but slightly more general with respect to loop formation) way to write the contribution to the SGWB per mode is~\cite{Auclair:2019wcv}
\beq
\Omega^{(k)}_{GW}(f) = \frac{1}{\rho_c}\frac{2k}{f}\,\Gamma^{(k)}G\mu^2
\int_{t_F}^{t_0}\!\!dt\;n(l_k(t),t)\,\lrf{a}{a_0}^5\Theta(t_i-t_F) \ ,
\label{eq:mode2}
\eeq
where $n(l,t)\,dl$ is the number density of loops with lengths between $l$ and $l+dl$ and
\beq
l_k(t) = \frac{2k}{f}\frac{a}{a_0}
\label{eq:lk}
\eeq
Matching this expression to the loop number density expected for $l(t_i) = \alpha t_i$ reproduces our previous expression.
\end{comment}
%%%%%%%%%%%%%%%%%%%%%%%

\subsection*{Analytic Estimates and Mode Sums}

As a practical matter, in computing the SGWB with the normal mode method only a finite number of modes can be kept . Previous analyses have argued that summing up to $k_{*} \sim 10^5$ total modes gives an excellent approximation of the cosmic string spectrum within the frequency ranges of interest to observation~\cite{Sanidas:2012ee,Sousa:2020sxs}. Using analytic estimates, we show that this guideline works well for string networks that reach scaling early on within a radiation dominated universe, but that it can fail for networks that approach scaling only relatively recently. In doing so, we also demonstrate that $\Omega_{GW}\propto f^{-1/3}$ is obtained from the normal mode method in the latter case when a sufficiently (and often very) large number of modes are included.

Consider the contribution to the SGWB at frequency $f$ from the lowest mode $k=1$ in the limit $t_F\to 0$. Assuming a standard cosmological evolution with radiation domination going back to very early times, for the frequency and parameter ranges of interest the integral of Eq.~\eqref{eq:mode1} for fixed frequency $f$ is typically dominated by emission at time $t \sim \bar{t}(f)$, defined by~\cite{Cui:2018rwi}
\beq
\frac{2}{f}\lrf{a(\bar{t})}{a_0} = \Gamma G\mu\,\bar{t} \ .
\label{eq:tbar}
\eeq
This time corresponds to the point at which the two terms in the expression for $t_i$ in Eq.~\eqref{eq:ti} are equal. For $t < \bar{t}$ the integrand of Eq.~\eqref{eq:mode1} increases more quickly than $t^{-1}$, while for $t > \bar{t}$ it decreases more rapidly than this. For a given $\bar{t}(f)$, there is a corresponding loop creation time given by 
\beq
t_i(\bar{t}(f)) = \frac{2\,\Gamma G\mu}{\alpha+\Gamma G \mu}\,\bar{t}
\ \ll \ \bar{t} \ .
\label{eq:tibar}
\eeq
Note as well that if $\bar{t}(f)$ occurs within a matter- or radiation-dominated era, it decreases monotonically with increasing frequency.

The characteristic feature of the SGWB from a scaling cosmic string network with $t_F\to 0$ is an approximately flat plateau at higher frequencies. This occurs for frequencies large enough that $\bar{t}(f) \lesssim t_{eq}$. Indeed, approximating the integral of Eq.~\eqref{eq:mode1} by the contributions near $t\sim \bar{t}(f)$ with $a(\bar{t})/a_0 \propto t^{1/2}$, one obtains
\beq
\Omega^{(1)}_W(f) \propto f^0 \ .
\eeq
The sum over higher normal modes can then be performed with the relation of Eq.~\eqref{eq:gwkscl}. As long as $\bar{t}(f/k)$ remains
less than $t_{eq}$, the relative contributions of these modes scale as
\beq
\Omega^{(k)}_{GW}(f) = k^{-4/3}\Omega^{(1)}_{GW}(f/k) 
\simeq k^{-4/3}\Omega^{(1)}_{GW}(f) 
\label{eq:lowtf}
\eeq
and the mode summation converges quickly enough that keeping only $k\leq k_* = 10^5$ modes is an excellent approximation.

Let us turn next to a diluted network of cosmic strings that only reach scaling relatively late in the history of the universe. A simple model for this that captures the essential features is to treat scaling (and loop emission) as beginning instantaneously at time $t_F$. We also take $t_F < t_{eq}$ since this is the most case of greatest interest. 

A non-zero formation time $t_F$ does not impact the resulting SGWB for frequencies $f$ such that $t_i(\bar{t}(f)) \gg t_F$. In this case, the dominant emission comes from loops formed after scaling is attained. This also applies to the contribution from the lowest normal mode, as well as those from higher modes since $t_i(\bar{t}(f/k)) \geq t_i(\bar{t}(f))$.

In contrast, the SGWB signal is reduced at larger frequencies where $t_i(\bar{t}(f)) < t_F$. When this occurs, the integration in Eq.~\eqref{eq:mode1} for $k=1$ is cut off by the step function before the dominant portion of the integrand is attained. The frequency dependence of the lowest mode is then
\beq
\Omega^{(1)}_W(f)  \propto  f^{-1} \ .
\eeq
This relation implies that higher normal modes are now more important than they would be for $t_F \to 0$. In particular, while we have $t_i(\bar{t}(f/k)) < t_F$, the higher modes contribute as
\beq 
\Omega^{(k)}_{GW}(f) = k^{-4/3}\Omega^{(1)}_{GW}(f/k) 
\simeq k^{-1/3}\Omega^{(1)}_{GW}(f) \ .
\label{eq:highfslope}
\eeq
The sum over these mode contributions would diverge if it continued indefinitely. Fortunately, as the mode number grows so too does the relevant loop formation time, and this provides a cutoff. Define $k_F$ by the relation
\beq
t_i(\bar{t}(f/k_F)) = t_F \ .
\label{eq:kf}
\eeq
When $k< k_F(f)$ the mode scaling of Eq.~\eqref{eq:highfslope} applies. However, when $k> k_F(f)$ we have $t_i(\bar{t}(f/k)) > t_F$ and 
\beq
\Omega^{(k)}_{GW}(f) 
\simeq k_F\,{k^{-4/3}}\,\Omega^{(1)}_{GW}(f) \ ,
\eeq
transitioning to a convergent sum on modes.

Evidently $k_* > k_F$ total modes must be included to obtain an accurate SGWB spectrum in the normal mode approach. For a given frequency $f$ and formation time $t_F$, the mode number $k_F$ can sometimes be much larger than the $k_{*} \sim 10^5$ that are typically kept in the normal mode method. Using the results above, we can estimate the effect of cutting off the mode sum at $k_{*} < k_F$ relative to the full result with $k_{*}\to \infty$. To do so, we approximate the sum on modes by an integral up to $k_{*}$,
\beq
\Omega_{GW}(f,k_*)  &=& \sum_{k=1}^{k_{*}}\Omega_{GW}^{(k)}(f)
\label{eq:scl}
\\ 
&\to&
\int_{1}^{k_{*}}\!dk\;\Omega_{GW}^{(k)}(f)
\nnmb\\
&\sim& \left\{
\begin{array}{ccc}
k_*^{2/3}\Omega_{GW}^{(1)}(f)&;&k_*\ll k_F\\
k_F^{2/3}\Omega_{GW}^{(1)}(f)&;&k_* \gg k_F
\end{array}\right.\nnmb
\eeq
As expected, $k_* \gtrsim k_F$ modes must be kept to get an accurate result. When $k_* \ll k_F$ the scaling with frequency follows that of $\Omega_{GW}^{(1)}(f) \propto f^{-1}$, which is the scaling relation found for diluted strings in Refs.~\cite{Guedes:2018afo,Gouttenoire:2019kij}. However, the full result with $k_*\to \infty$ typically has a different frequency scaling because $k_F$ can also depend on $f$. In particular, solving Eqs.~(\ref{eq:ti},\ref{eq:tibar},\ref{eq:kf}) for $t_F < t_{eq}$ with the approximation $(a/a_0) \simeq z_{eq}^{-1/4}(t/t_0)^{1/2}$ (and $z_{eq} \simeq 3390$) gives
\beq
k_F \ \simeq \ \left(\frac{\alpha}{8}z_{eq}^{1/2}\Gamma G\mu\right)^{1/2} (t_0t_F)^{1/2}\,f \ .
\eeq
Combining this with the estimate of Eq.~\eqref{eq:scl}, we find $\Omega_{GW} \propto f^{-1/3}$ at high frequencies, matching the scaling we obtained using the burst method.

The necessity of including a very large number of modes to compute the SGWB in the normal mode approach applies beyond the specific scenario of a diluted cosmic string network with a relatively large scaling time $t_F$. In particular, our analysis generalizes to a range of scenarios where the cosmic string network reaches scaling early on ($t_F \to 0$) but whose SGWB spectrum is modified by a period of non-minimal cosmological evolution that transitions to the standard radiation domination at time $t_{\Delta}$~\cite{Cui:2017ufi,Cui:2018rwi,Gouttenoire:2019kij,Gouttenoire:2019rtn}. Within many of these scenarios, including most notably an early period of matter domination, the contribution from the lowest normal mode is found to go like $\Omega_{GW}^{(1)} \propto f^{-1}$ when $\bar{t}(f) < t_{\Delta}$~\cite{Cui:2018rwi}. The approach above carries over directly to these scenarios with the simple replacement $t_F\to t_{\Delta}$. We note that this implies further that the fall off of the SGWB at high frequencies from an early period of matter domination goes as $\Omega_{GW} \propto f^{-1/3}$, and not as $\Omega_{GW} \propto f^{-1}$ quoted in Refs.~\cite{Cui:2017ufi,Cui:2018rwi,Gouttenoire:2019kij,Gouttenoire:2019rtn}.

\subsection*{A Useful Approximation}

 The analysis above shows that a sufficient number of modes must be included for the normal mode approach to give the correct SGWB. In practice, summing over very large numbers of modes becomes computationally expensive. A simple approach to include an arbitrarily high number of modes in an accurate and efficient way is to sum up to $k=N$ modes discretely and approximate the sum over all higher modes by an integral,
 \beq
 \Omega_{GW}(f) \simeq \sum_{k=1}^N\Omega_{GW}^{(k)}
 + \int_{{N+1}}^{\infty}\!dk\;\Omega_{GW}^{(k)} \ .
 \eeq
 This approximation is typically accurate up to corrections on the order of $1/N$.
 
 \subsection*{Emission from kinks and other sources}
 
 In this supplement we focused on gravitational wave emission from cusps since these are typically assumed to be the dominant source~\cite{Blanco-Pillado:2017oxo}. However, the discussion above can be generalised to other localized sources of GWs from cosmic strings, beyond just cusps. These can also be treated in the normal mode method, but with a different scaling index $q$ on the relative power per mode that generalizes Eq.~\eqref{eq:pmode} to 
\beq
\Gamma = \sum_k\Gamma^{(k)} \ ,~~~~~
\Gamma^{(k)} \simeq \Gamma \frac{k^{-q}}{\sum_{m=1}^{\infty} m^{-q}} \ .
\eeq
Taking $q = 4/3$ reproduces Eq.~\eqref{eq:pmode} for cusps, while $q=5/3$ corresponds to the emission from cosmic string kinks~\cite{Damour:2001bk}. For a given $q$, Eq.~\eqref{eq:gwkscl} is also modified to
\beq
\Omega_{GW}^{(k)} = k^{-q}\Omega_{GW}^{(1)}(f/k) \ .
\eeq
Generalizing the analysis above to general $q > 1$, the slope at high frequencies with late scaling or early matter domination becomes $\Omega_{GW} \propto f^{1-q}$.

 \bigskip
 
 \textbf{Note Added:} While this material was being prepared, a similar observation and approximation for cosmic string mode summations was made in Ref.~\cite{Blasi:2020wpy} (as well as a version 2 of Ref.~\cite{Gouttenoire:2019kij}). The authors of Ref.~\cite{Blasi:2020wpy} also demonstrated the importance of keeping a very large number of normal modes when computing the SGWB from a scaling cosmic string network with an early period of cosmological matter domination. As we showed above, this corrects the high frequency scaling relations quoted in Refs.~\cite{Cui:2017ufi,Cui:2018rwi,Gouttenoire:2019kij,Gouttenoire:2019rtn}.
%%%%%%%%%%%%%%%%%%%%%%%%%%%%%%%%%%%%%%%%%%%%%%%%%%%%%%%%%%%%%%%%%%%%%%%%%%%%%%%%%%%%%%%%%%%%%%%%%%%%
%%%%%%%%%%%%%%%%%%%%%%%%%%%%%%%%%%%%%%%%%%%%%%%%%%%%%%%%%%%%%%%%%%%%%%%%%%%%%%%%%%%%%%%%%%%%%%%%%%%%

\bibliography{csregrowth}
%%%%%%%%%%%%%%%%%%%%%%%%%%%%%%%%%%%%%

\end{document}